\UseRawInputEncoding
\documentclass[%
 pra,
 mph,%
 amsmath,amssymb,
 reprint,%
]{revtex4-2}

\usepackage{graphicx}
\usepackage{dcolumn}
\usepackage{bm}
\usepackage{hyperref}
\usepackage{float}
\hypersetup{
    hypertex=true,
    colorlinks=true,
    linkcolor=blue,
    filecolor=blue, 
    citecolor=blue,     
    urlcolor=blue,
    anchorcolor=blue}

\usepackage[mathlines]{lineno}

\begin{document}

\preprint{AAPM/123-QED}

\title{Rotating dipole and quadrupole quantum droplets in binary Bose-Einstein condensates}
\author{Dongshuai Liu$^1$}
\author{Yanxia Gao$^2$}
\author{Dianyuan Fan$^1$}
\author{Boris A. Malomed$^3$} 
\thanks{Sabbatical address}

\author{Lifu Zhang$^1$}%
\email{zhanglifu@szu.edu.cn.}
\affiliation{$^1$International Collaborative Laboratory of 2D Materials for Optoelectronic Science $\&$ Technology, Institute of Microscale Optoelectronic, Shenzhen University, Shenzhen, 518060, China}
\affiliation{$^2$School of Physics and Optoelectronic Engineering, Shenzhen University, Shenzhen, 518060, China}
\affiliation{$^3$Instituto de Alta Investigaci\'{o}n, Universidad de Tarapac\'{a}, Casilla 7D, Arica, Chile}

\date{\today}

\begin{abstract}

	Quantum droplets (QDs) are self-trapped modes stabilized by the Lee-Huang-Yang correction to the mean-field Hamiltonian of binary atomic Bose-Einstein condensates. The existence and stability of quiescent and rotating dipole-shaped and vortex QDs with vorticity $S=1$ (DQDs and VQDs, respectively) are numerically studied in the framework of the accordingly modified two-component system. The rotating DQDs trapped in an annular potential are built of two crescent-like components, stretching along the azimuthal direction with the increase of the rotation frequency. Rotating quadrupole QDs (QQDs) bifurcate from the VQDs with $S=2$. Above a certain rotation frequency, they transform back into VQDs with a flat-top shape. Rotating DQDs and QQDs are stable in a broad interval of values of the chemical potential. The results provide the first example of stable modes which are intermediate states between the rotating DQDs and QQDs on the one hand, and VQDs on the other.
\end{abstract}

\maketitle

\section{INTRODUCTION}

Droplets of ultradilute quantum liquids represent a new quantum state of matter in the realm of Bose-Einstein condensates (BECs) \cite{cuispinorbitcouplinginduced2018, luo_new_2020}. The formation of the quantum droplets (QDs) is provided by the balance between the effective mean-field (MF) attraction, which, in turn, is a result of the competition of interspecies attraction and intraspecies repulsion, and beyond-MF
self-repulsion in each component, induced by quantum fluctuations around the MF states, which is represented by the Lee-Huang-Yang (LHY) corrections to the MF theory \cite{lee_eigenvalues_1957}. In terms of the two-dimensional (2D) Gross-Pitaevskii equations (GPEs), the LHY effect amounts to the logarithmic factor multiplying the usual cubic term, while in the 3D and 1D settings the LHY corrections are represented, respectively, by the additional self-repulsive quartic and attractive quadratic terms in the respective GPEs. QDs have been predicted theoretically \cite{petrov_quantum_2015,petrov_ultradilute_2016} and created experimentally in dipolar bosonic gases \cite{chomaz_quantum-fluctuation-driven_2016,
schmitt_self-bound_2016}, as well as in mixtures of two atomic states with contact interactions \cite%
{cabrera_quantum_2018,cheiney_bright_2018,semeghini_self-bound_2018,derrico_observation_2019}. Typically, QDs are composed of several thousands of atoms. They are droplets of extremely dilute quantum fluids, whose densities are more than eight orders of magnitude lower than in liquid helium, which is the
\textquotedblleft exemplary\ "quantum fluid. Featuring robust inner coherence and being well controlled by experimental parameters, QDs offer unique advantages for the implementation of quantum simulations and
metrology \cite{clark_universal_2016, leonard_supersolid_2017, liang_laser_2020}.

Internal superfluidity of QDs suggests that they may maintain vortex states, which are characterized by the respective topological charge (winding number) and zero density at the pivot \cite{fetter_rotating_2009}. The vorticity may be imprinted onto QDs by an appropriate phase structure in the initial state.

{In the simplest case, which was is adopted in a majority of theoretical works on QDs, the balanced binary BEC, whose components have identical shapes, can be described by a single wave function, obeying the single GPE, provided that scattering lengths of inter-atomic collisions are equal in the two components. The latter condition can be readily maintained by the Feshbach resonance in the two hyperfine states. The analysis has demonstrated that the single-component reduction of the full system of two coupled GPEs is stable against small perturbations which break the equality of the two components \cite{li_two-dimensional_2018}. A special case is the two-component state with {\it hidden vorticity}, when the two components assume the vortical shape with identical amplitude profiles but opposite winding numbers (topological charges). In the latter case, the full two-component system should be used, the result being that such two-component ``hidden-vorticity" states are chiefly unstable, but, nevertheless, they feature a small stability region in the respective parameter manifold. In the case of the effectively 2D QDs with imbalanced components (carrying different numbers of atoms), trapped in a confining potential, the full system should be used too \cite{flynn_harmonically_2024}. The analysis demonstrates that, in the limit of the balanced binary BEC (which is the case addressed in the present work), the results produced by the two-component system smoothly carry over into those predicted  by the single GPE for identically equal components.

Moreover, if the inter-atomic attraction is provided by the long-range dipole-dipole interaction, a single-component condensate composed of dipolar atoms is sufﬁcient for the creation of QDs \cite{schmitt_self-bound_2016, ferrier-barbut_observation_2016}.

Stable 2D and 3D vortex QDs (VQDs) have been predicted in the binary BEC with contact inter-atomic interactions \cite{li_two-dimensional_2018, kartashov_three-dimensional_2018,huang_binary-vortex_2022,
dong_internal_2022, dong_stable_2023}, while vortex states embedded in dipolar QDs are unstable against spontaneous fragmentation \cite{cidrim_vortices_2018, lee_excitations_2018}. Ring-shaped QDs may also support semidiscrete \cite{zhang_semidiscrete_2019} and higher-order vortex structures \cite{huang_stable_2023, dong_stable_2024}. Vortex clusters can be generated too, as the ground state of rotating trapped binary BECs \cite{tengstrand_rotating_2019, jiang_vortex_2022}. Experimental evidence of VQDs has not been reported as yet, which suggests looking for new settings that may be conducive to the existence of such stable topological modes.

Other vortex species of stable QDs have been predicted in the form of metastable necklace-shaped structures carrying angular momentum \cite{kartashov_metastability_2019,kartashov_multicolor_2002}, including two-component necklace patterns \cite{desyatnikov_necklace-ring_2001, kartashov_robust_2002}. Solutions for stable rotating QDs with whispering-gallery-like shapes were found under the action of broad 2D trapping potentials \cite{dong_rotating_2021}. Bistable multipole QDs were predicted in symmetric binary BECs \cite{dong_bistable_2022}. QD crystals were shown to exist in an axially symmetric harmonic-oscillator (HO) trapping potential \cite{baillie_droplet_2018}. QDs with heterosymmetric and heteromultipole structures may also be stable \cite{kartashov_structured_2020}.

Another possibility for trapping nonlinear modes is offered by the use of optical-lattice (OL) potentials \cite{brazhnyi_theory_2004,morsch_dynamics_2006}. The balance between the intercomponent attraction, repulsive LHY correction, and the OL trapping effect provides for the existence of stable QDs under broad conditions \cite{morera_universal_2021,zhou_dynamics_2019, morera_quantum_2020,zheng_quantum_2020, liu_vortex-ring_2022}. In this context, the dynamics of QDs with mutually symmetric spinor components was studied in the presence of the OL potential \cite{dong_multi-stable_2020}. On-site- and intersite-centered semidiscrete QDs were predicted in arrays of nearly-1D traps \cite{zhang_semidiscrete_2019}. Further, 1D multihumped QDs were explored under the action of spatially-periodic modulations of the nonlinearity \cite{chen_one-dimensional_2021}. These findings reveal that OL potentials provide a versatile platform for the study of QDs.

Although the creation of various QDs configurations has been predicted, intermediate states, which bridge rotating dipole-shaped QDs (DQDs), that are built as a bound state of two oppositely placed crescent-like fragments, and quadrupole QDs (QQDs) to VQDs, were not addressed previously. This is the subject of the present work. While rotating DQDs and QQDs are obviously unstable in the free space, they may be stabilized by a combined HO-Gaussian annular potential. In particular, we demonstrate that, in the presence of the trapping annular potential, DQDs and QQDs transform into VQDs with the increase of the rotation velocity. The rotating DQDs and QQDs, along with VQDs, are robust modes in a broad interval of values of the corresponding chemical potential.

\section{THEORETICAL MODEL}

In the 2D setting, we consider the evolution of the two-component MF wave function $\psi _{\pm }(x,y,t)$ of the symmetric binary BEC, assuming, as said above, that the inter- and intra-species contact interactions are
attractive and repulsive, respectively. The respective scaled system of coupled GPEs is \cite{li_two-dimensional_2017}

\begin{equation}  \label{eq:refname1}
\begin{split}
{i}\frac{\partial \psi _{\pm }}{\partial {t}}= & -\frac{1}{2}\left( \frac{
\partial ^{2}\psi _{\pm }}{\partial x^{2}}+\frac{\partial ^{2}\psi _{\pm }}{
\partial y^{2}}\right) +\frac{4\pi }{g}(|\psi _{\pm }|^{2}-|\psi _{\mp
}|^{2})\psi _{\pm } \\
& +(|\psi _{+}|^{2}+|\psi _{-}|^{2})\ln (|\psi _{+}|^{2}+|\psi
_{-}|^{2})\cdot \psi _{\pm },
\end{split}%
\end{equation}

\noindent where the logarithmic factor represents the LHY modification of the MF nonlinearity. Here, the wave functions $\psi _{\pm }$, coordinates $(x,y)$, and time $t$ are measured in units of $\sqrt{n_{0}}$, $g/2\sqrt{\pi } $, and $g^{2}/4\pi $, respectively, where $n_{0}$ is the equilibrium density \cite{petrov_ultradilute_2016}, and $g>0$ is the coupling constant. For this system, we consider the most natural symmetric bound states, with $\psi _{+}=\psi _{-}\equiv \psi /\sqrt{2}$. To address rotating QDs, we introduce the rotating coordinate frame with angular velocity $\omega $, $x^{^{\prime }}=x\cos (\omega t)+y\sin (\omega t)$, $y^{^{\prime }}=y\sin (\omega t)-x\cos (\omega t)$. Adding a confining axisymmetric potential $V(r)$, where $r$ is the radial coordinate, the corresponding single GPE is written, in the rotating coordinates, as

\begin{equation}
{i}\frac{\partial \psi }{\partial {t}}=-\frac{1}{2}\nabla ^{2}\psi +|\psi
|^{2}\ln (|\psi |^{2})\cdot \psi +V(r)\psi -\omega L_{z}\psi \mathrm{,}
\label{eq:refname2}
\end{equation}

\noindent where $L_{z}=xp_{y}-yp_{x}$ is the angular-momentum operator. The confining potential is taken here as

\begin{equation}
{V(r)}=\frac{1}{2}\Omega ^{2}r^{2}+V_{0}\exp ({-r^{2}}/{a}^{2}),  \label{V}
\end{equation}

\noindent with $V_{0}>0$. It is a combination of the HO trap and Gaussian potential hill at the center. Fixing $\Omega =0.1$ for the shallow HO trap by means of scaling, generic numerical results are presented below for the Gaussian amplitude and width $V_{0}=0.4$ and $a=5$, cf. Refs. \cite{dong_rotating_2022, liu_multi-stable_2023}. Accordingly, the first and second terms in potential (\ref{V}) dominate at $r>5$ and $r<5$, respectively. Furthermore, for these parameters, the second (Gaussian) term in Eq. (\ref{V}) dominates over the other repulsive potential, \textit{viz}., the vorticity-induced one, $S^{2}/\left( 2r^{2}\right) $, for vortex states with integer winding number $S$, in the interval of $1<r<10$, for $S=1 $. This conclusion implies that the results reported below are essentially determined by the Gaussian term.

It is relevant to mention that the stability of VQDs with high values of the winding number (at least, up to $S=12$) was recently investigated in a similar model, with the same nonlinearity as in Eq. (\ref{eq:refname2}) and an annular potential which, unlike one (\ref{V}), is a Gaussian-shaped trough, which does not include the HO term \cite{dong_stable_higher_charge_2023}.

In the case of VQDs, Eq. (\ref{eq:refname2}) can be rewritten, in the polar coordinates $\left( r,\theta \right) $, as

\begin{equation}
{i}\frac{\partial \psi }{\partial {t}}-i\omega \frac{\partial \psi }{
\partial {\theta }}=-\frac{1}{2}\nabla ^{2}\psi +|\psi |^{2}\ln (|\psi
|^{2})\cdot \psi +V(r)\psi .  \label{Omega}
\end{equation}

\noindent Bound state produced by Eq. (\ref{eq:refname2}) are characterized by the norm,

\begin{equation}
N=\iint {|\psi |^{2}dxdy}.  \label{N}
\end{equation}

\noindent Then stationary solutions to Eq. (\ref{Omega}) for VQDs with integer vorticity $S$ are looked for as

\begin{equation}
\psi =\exp \left( -i\mu t+iS\theta \right) U(r),  \label{psi}
\end{equation}

\noindent where real function $U(r)$ obeys the radial equation

\begin{equation}
\begin{split}
\left( \mu +\omega S\right) U=& -\frac{1}{2}\left( \frac{d^{2}U}{dr^{2}}+%
\frac{ 1}{r}\frac{dU}{dr}-\frac{S^{2}}{r^{2}}U\right) \\
& +2U^{3}\ln (U)+V(r)U.
\end{split}
\label{U}
\end{equation}

\noindent Thus, for the rotating VQDs, the rotation effect amounts to the shift of the chemical potential, $\mu \rightarrow \mu +\omega S$. In the absence of the rotation, stability of VQDs in the present model was
investigated in Ref. \cite{li_two-dimensional_2018}.

Stability of QDs is addressed below by considering perturbed solutions,

\begin{equation}
\begin{split}
\psi ({x},{y},{z})&=\left[ U (x,y)+{u(x,y)\exp }\left( \lambda {t}%
\right)\right. \\
&+ \left.{\ \ v^{\ast }(x,y)\exp }\left( \lambda ^{\ast }{t}\right) \right] {%
\exp }(-{i}{\ \ \mu }{t}),
\end{split}
\label{eq:refname4}
\end{equation}

\noindent where ${u}\left( {x,y}\right) $ and ${v}\left( {x,y}\right) $ are eigenmodes of infinitesimal perturbations, $\lambda $ is the corresponding growth rate, and $\ast $ stands for the complex conjugate. The substitution of the perturbed wave form (\ref{eq:refname4}) in Eq. (\ref{eq:refname2}) and linearization leads to the eigenvalue problem for $\lambda $, based on the respective Bogoliubov -- de Gennes (BdG) equations:

\begin{equation}
{i}{\ \left(
\begin{array}{cc}
{F}_{11} & {F}_{12} \\
-{F}_{12}^{\ast } & -{F}_{11}^{\ast }%
\end{array}%
\right) \left(
\begin{array}{c}
{u} \\
{v}%
\end{array}%
\right) =\lambda \left(
\begin{array}{c}
{u} \\
{v}%
\end{array}%
\right) ,}  \label{eq:refname5}
\end{equation}

\noindent with ${F}_{11}\equiv -\frac{1}{2}\nabla ^{2}-{\mu }+{V}+2|U|^{2}\left[ \ln (\lvert U\rvert ^{2}+\frac{1}{2})\right] -i\omega \left( x\frac{\partial }{\partial y}-y\frac{\partial }{\partial x}\right) $ and $F_{12}=U^2 \left[ \ln (\lvert U \rvert ^{2}+1)\right]$.
The stationary QD is stable if all eigenvalues $\lambda $ are imaginary.

To produce stationary QDs solutions of Eq. (\ref{eq:refname2}), the Newton's iterative method was used. Their stability was identified as per the spectrum of eigenvalues $\lambda $, provided by the numerical solution of BdG equations (\ref{eq:refname5}), which was performed by means of the Fourier collocation method, and verified in simulations of the perturbed QD evolution, performed by means of the split-step fast-Fourier-transformed method.

\section{NUMERICAL RESULTS AND DISCUSSIONS}

Typical profiles of the DQDs and VQDs with $S=1$ are plotted in Fig.~\ref{fig:1}, and families of such states are presented by means of the respective $N(\mu )$ curves in Fig. \ref{fig:2}(a,b). In the absence of
rotation, i.e., for $\omega =0$, there exist two branches of the $N(\mu )$ dependences with slopes of opposite signs. DQDs belonging to the upper branch are broader, being built of crescent-shaped lobes. It is seen in Figs. \ref{fig:1}(b1-b3) that, as $\omega $ grows, the lobes gradually expand along the azimuthal direction, and eventually fuse into a VQD with $S=1$ at a critical angular velocity.

\begin{figure}[tbph]
\centering
\includegraphics[width=\linewidth]{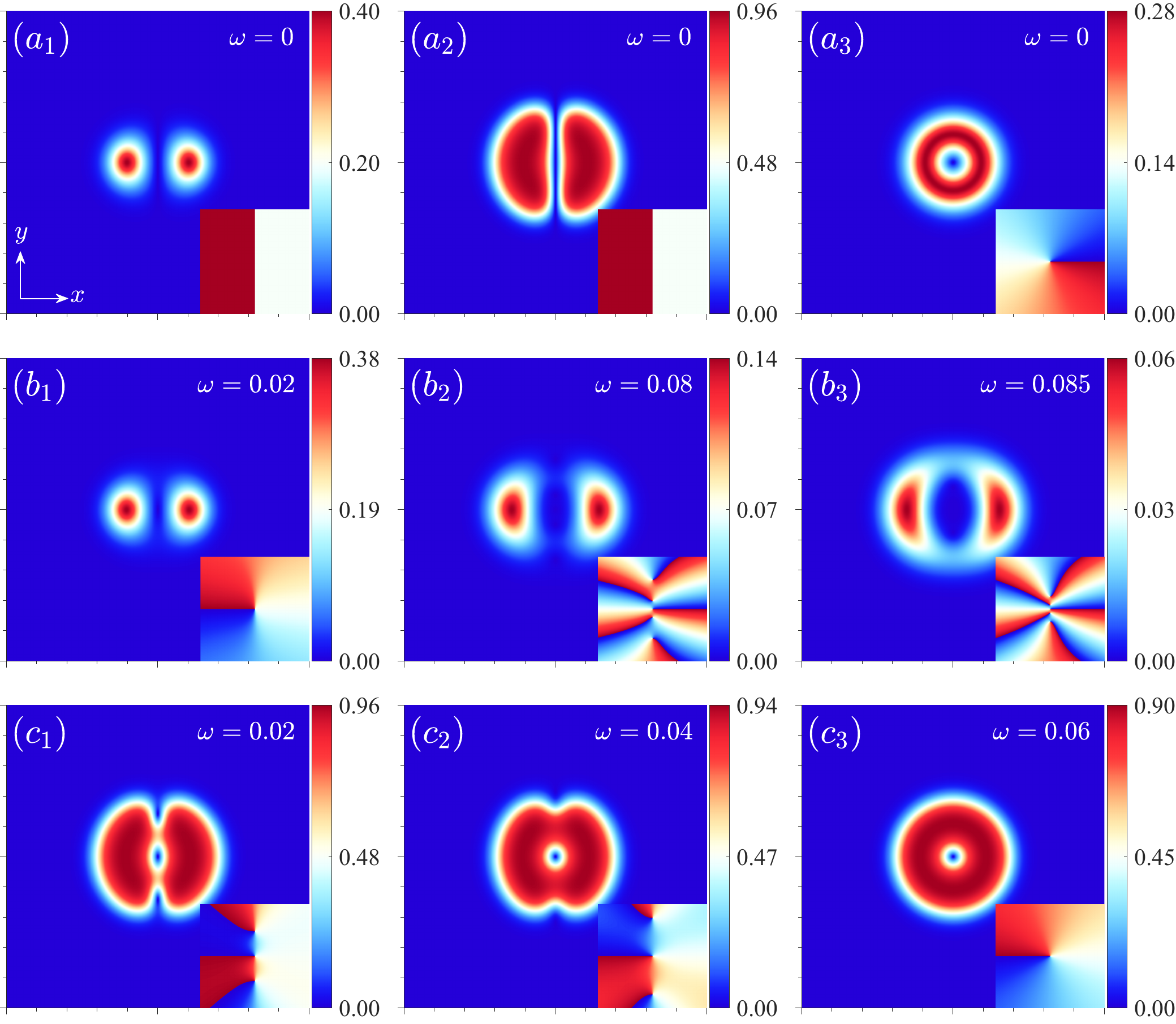} 
\caption{$(a_{1})-(a_{3})$ The absolute value of the field in the dipole-shaped and vortex quantum droplets (DQDs and VQDs) at $\protect\omega =0$, which are marked in Fig.~\protect\ref{fig:2}(a). DQDs shown in $(a_{1})$ and $(a_{2})$ belong to the lower and upper branches, respectively. $(a_{3})$ A lower-branch VQD with $S=1$. $(b_{1})-(b_{3})$ Lower-branch DQDs marked in Fig.~\protect\ref{fig:2}(c), for $\protect\omega %
=0.02$, $\protect\omega =0.08$, and $\protect\omega =0.085$, respectively. $(c_{1})-(c_{3})$ Upper-branch dipole droplets marked in Fig.~\protect\ref{fig:2}(d) at $\protect\omega =0.02$, $\protect\omega =0.04$, and $\protect\omega =0.06$, respectively. All panels pertain to $\protect\mu =0.21$. This and similar figures below display the solutions in the domain $(x,y)\in [-25,+25]$.}
\label{fig:1}
\end{figure}

In Figs.~\ref{fig:2}(a,b) the lower and upper branches $N(\mu )$ for DQDs with $\omega =0$ monotonously decreases and increases, respectively, with the growth of $\mu $. They merge at the lower cutoff value of the chemical potential. When $\omega =0.05$, the $N(\mu )$ curve of DQDs bifurcates from the VQD with $S=1$ and merges with the curve for the VQD family at a lower value of $N$. In Fig.~\ref{fig:2}(c), at fixed $\mu $ the norm of the lower-branch rotating DQD monotonously decreases with the increase of the angular velocity $\omega $. The families of the rotating DQDs and VQDs merge at $\omega $ reaching its maximum value. In other words, the DQDs originate from the vortex eigenmode of the rotating linear system. Accordingly, the limit values of $\mu$ and $\omega$ for $N\to 0$ in Figs. \ref{fig:2}(a,b) and (c), respectively, correspond to the eigenvalues of the solution of the linear Schr\"{o}dinger equation, which is the linear limit of Eq. (\ref{U}). On the other hand, in Fig.~\ref{fig:2}(d) the norm of the upper-branch DQDs is a nonmonotonous function of $\omega $. First, it increases and reaches a maximum, and then gradually decreases. In the course of this evolution, the DQD carries over into the VQD.

\begin{figure}[tbph]
\centering
\includegraphics[width=\linewidth]{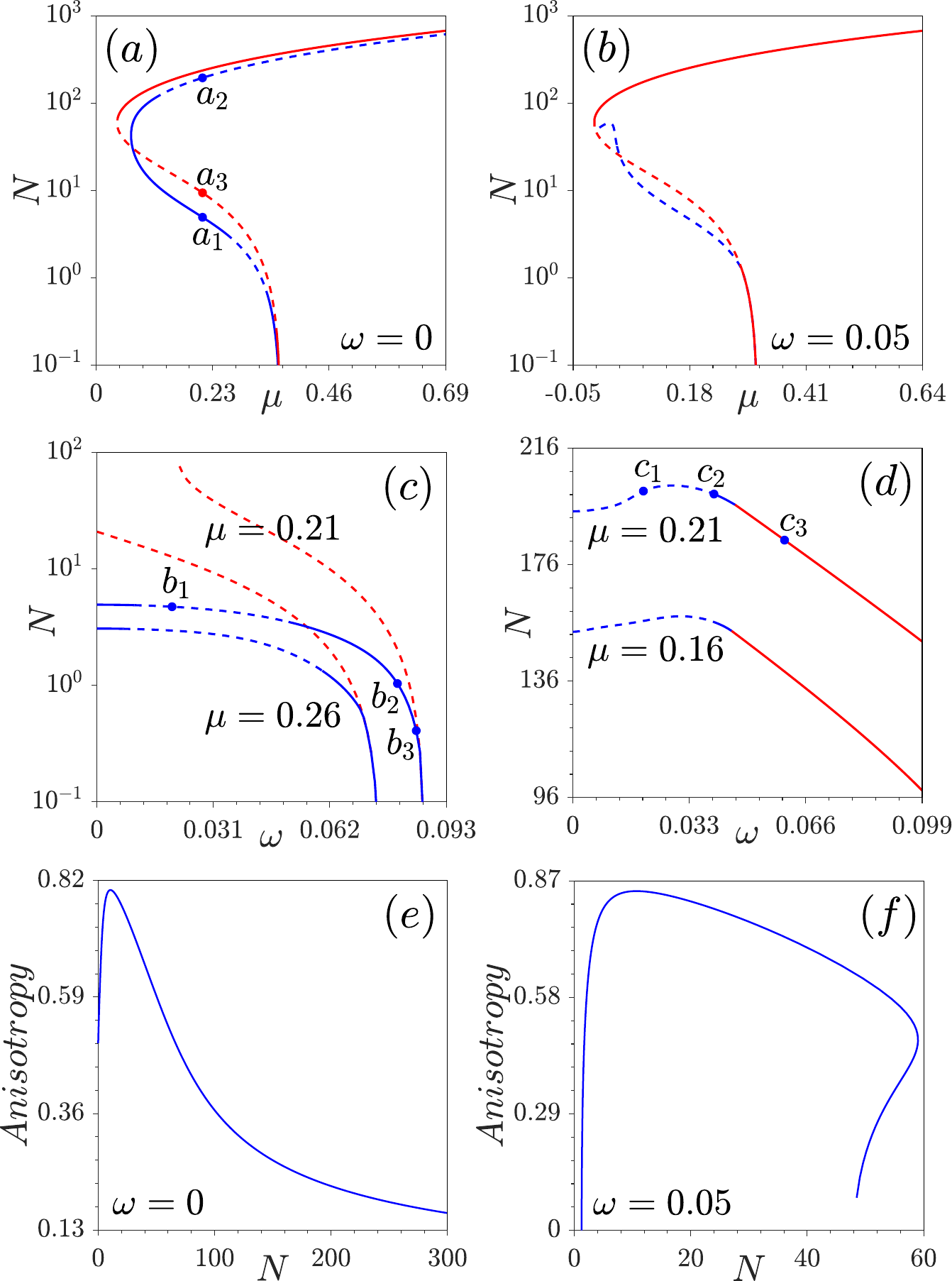} 
\caption{Norm $N$ versus $\protect\mu $ for DQDs (blue) and VQDs (red) at
(a) $\protect\omega =0$ and (b) $\protect\omega =0.05$. (c) $N$ vs. $\protect%
\omega $ for the lower-branch DQD family (blue) and VQD one (red) at $%
\protect\mu =0.21$ and $\protect\mu =0.26$. (d) $N(\protect\omega )$ curves
for the upper-branch DQD family at $\protect\mu =0.16$ and $\protect\mu %
=0.21 $. In this panel, the red lines represent the VQDs emerging from the
DQD states. Dashed and stable segments represent unstable and stable
subfamilies, respectively. (e,f) Anisotropy (\protect\ref{A}) of the DQD
density pattern vs. norm $N$ at (e) $\protect\omega =0$ and (f) $\protect%
\omega =0.05$. }
\label{fig:2}
\end{figure}

It is relevant to introduce a parameter characterizing the azimuthal anisotropy of the density pattern, $\left\vert \psi \left( r,\theta \right) \right\vert ^{2}$, of the stationary DQD states:

\begin{equation}
\mathrm{Anisotropy}=\frac{\int_{0}^{2\pi }rdr\int_{0}^{2\pi }\cos \left(
2\theta \right) d\theta \left\vert \psi \left( r,\theta \right) \right\vert
^{2}}{\int_{0}^{2\pi }rdr\int_{0}^{2\pi }d\theta \left\vert \psi \left(
r,\theta \right) \right\vert ^{2}}.  \label{A}
\end{equation}

\noindent The dependence of $\mathrm{Anisotropy}$ on norm $N$ is displayed in Figs.~\ref{fig:2}(e,f). At $\omega =0$, the $\mathrm{Anisotropy}$ of DQDs has inflexion points corresponding to the droplets expanding with the growth of $N$. At $\omega =0.05$, the $\mathrm{Anisotropy}$ starts from zero at the point where the DQD bifurcates from the VQD, as shown in Fig.~\ref{fig:2}(f).

The annular potential defined as per Eq. (\ref{V}) is crucially important for the stabilization of the rotating DQDs against splitting into fragments, as well as against decay towards $r\rightarrow \infty $. Results of the stability analysis results are presented in Fig.~\ref{fig:3} [note that the
Vakhitov-Kolokolov criterion, $dN/d\mu <0$ \cite{vakhitov_stationary_1973,berge_wave_1998}, cannot predict the stability in the present case, as the nonlinearity in Eq. (\ref{eq:refname2}) changes its sign with the increase of the density]. In particular, the instability growth rate $\lambda _{\mathrm{re}}$ for DQDs at $\omega =0$, produced by the numerical solution of the BdG equations (\ref{eq:refname5}), is displayed in Fig.~\ref{fig:3}(a). It is seen that the DQDs are stable in a large domain in the presence of the HO-Gaussian annular potential, similar to the case of a weakly anharmonic trapping potential \cite{dong_bistable_2022}. The lower-branch VQDs are unstable in their almost whole existence domain [Fig.~\ref{fig:3}(b)], while the upper-branch is completely stable. This is different from the case of the lower-branch
vortices with $S=1$ in the HO trapping potential, where there is a stability domain \cite{liu_higher-charged_2023}. The instability of the VQD may be considered as the modulational instability (MI) against azimuthal perturbations, which break the vortex' axial symmetry (generally, MI leads to self-induced breakup of initially homogenous waves in nonlinear media). \cite{ye_suppression_2004}
The rotating DQDs show a large stability domain. We display the instability-growth-rate curves for the DQDs with $\mu =0.21$ in Figs.~\ref{fig:3}(c,d). The lower and upper branches of DQDs feature an bistability
area at $\omega \in [ 0.063,+0.086]$.

\begin{figure}[tbph]
\centering
\includegraphics[width=\linewidth]{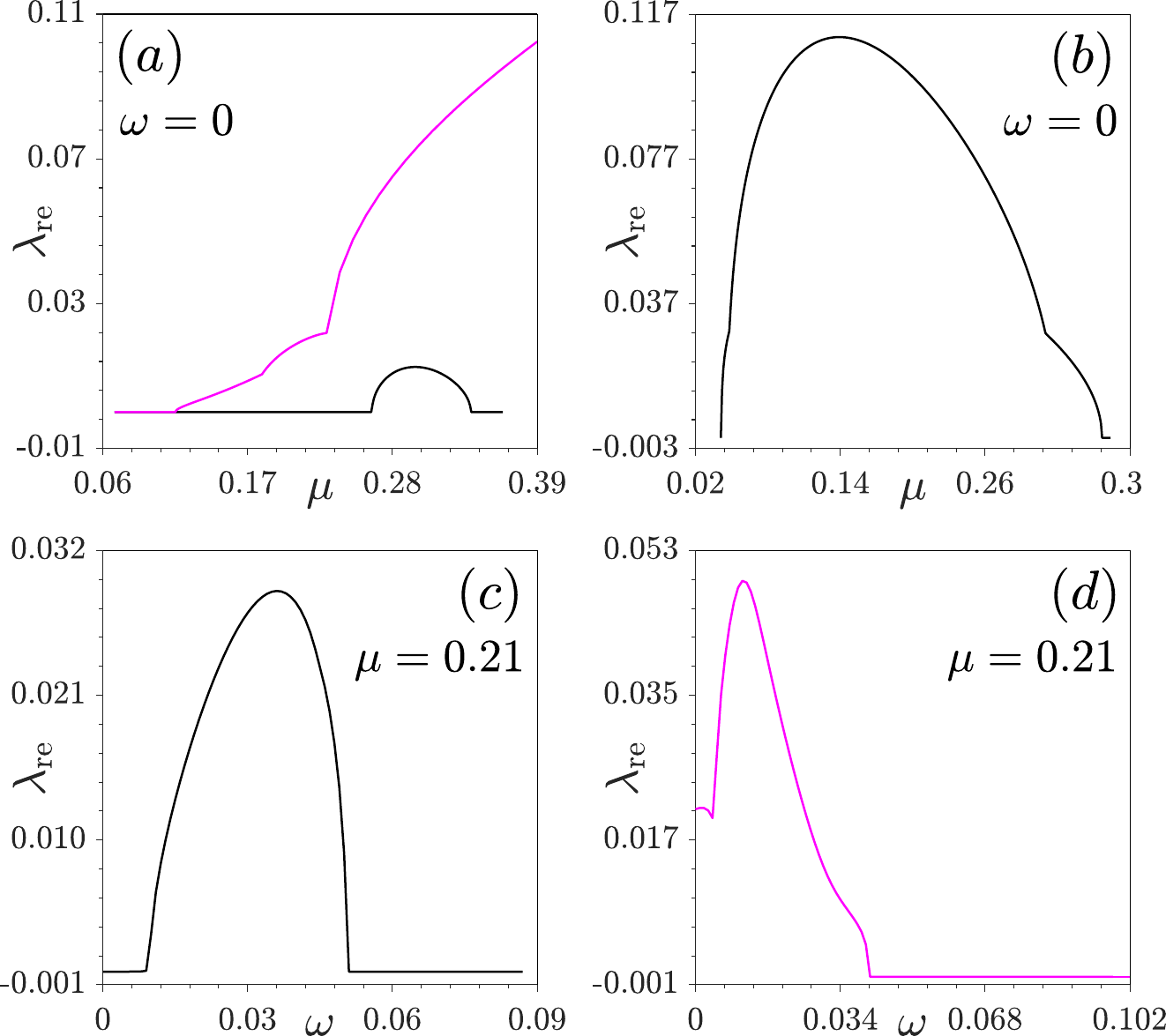}
\caption{Instability growth rate $\protect\lambda $ versus $\protect\mu $,
as obtained from the numerical solution of the BdG equations (\protect\ref%
{eq:refname5}) for (a) DQDs and (b) lower-branch VQDs. The instability
growth rate vs. $\protect\omega $ for (c) lower-branch and (d) upper-branch
DQDs at $\protect\mu =0.21$. The results for the lower and upper branches
are represented by the black and magenta curves, respectively. }
\label{fig:3}
\end{figure}

Predictions of the stability analysis based on the BdG equations (\ref{eq:refname5}) have been verified by systematically performed direct simulations of the perturbed evolution of the droplets. Typical examples of
the evolution are exhibited in Fig.~\ref{fig:4}. First, we test the DQDs and VQDs at $\omega =0$, as shown in Fig.~\ref{fig:4}(a). It is observed that even unstable DQDs survive for a long time, see Fig.~\ref{fig:4}($a_{2}$). Stable rotation of the DQDs is illustrated in Figs. \ref{fig:4}(b,c) by snapshots of the profiles of the absolute value of the wave function at different moments of time, with respect to rotation period $2\pi /\omega $.

\begin{figure}[tbph]
\centering
\includegraphics[width=\linewidth]{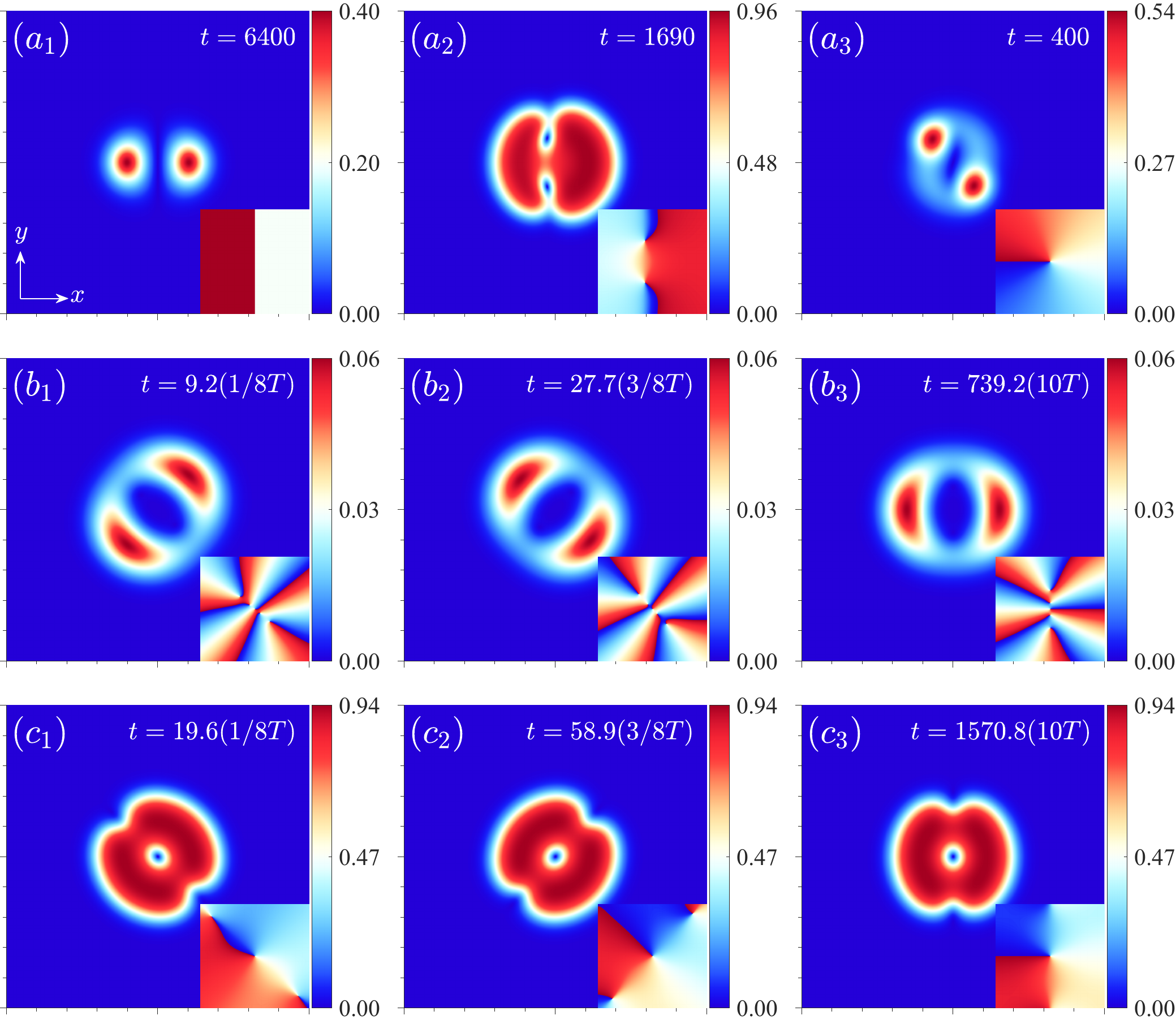} 
\caption{($a_{1}$) Stable evolution of a lower-branch DQD. ($a_{2}$) and ($%
a_{3}$): Unstable evolution of an upper-branch DQD and VQD, respectively.
(b) and (c): Stable evolution of lower- and upper-branch DQDs, respectively.
The rotation frequency is $\protect\omega =0$ in ($a_{1}-a_{3}$), $\protect%
\omega =0.085$ in ($b_{1}-b_{3}$), and $\protect\omega =0.04$ in ($%
c_{1}-c_{3}$). The chemical potential is fixed as $\protect\mu =0.21$ in all
panels. }
\label{fig:4}
\end{figure}

Next, we address the existence, stability, and evolution dynamics of rotating QQDs and VQDs with $S=2$. Representative shapes of these states are displayed in Fig.~\ref{fig:5}, including the QQDs with smaller and larger norms [Figs.~\ref{fig:5}($a_{1}$) and ($a_{2}$)]. Similar to the DQDs and VQDs with $S=1$, the droplets belonging to the upper branch are broader than on the lower branch. The trend for the transformation of the QQDs into VQDs with the increase of the rotation frequency $\omega $ is displayed in Figs.~\ref{fig:5}(b,c).

\begin{figure}[tbph]
\centering
\includegraphics[width=\linewidth]{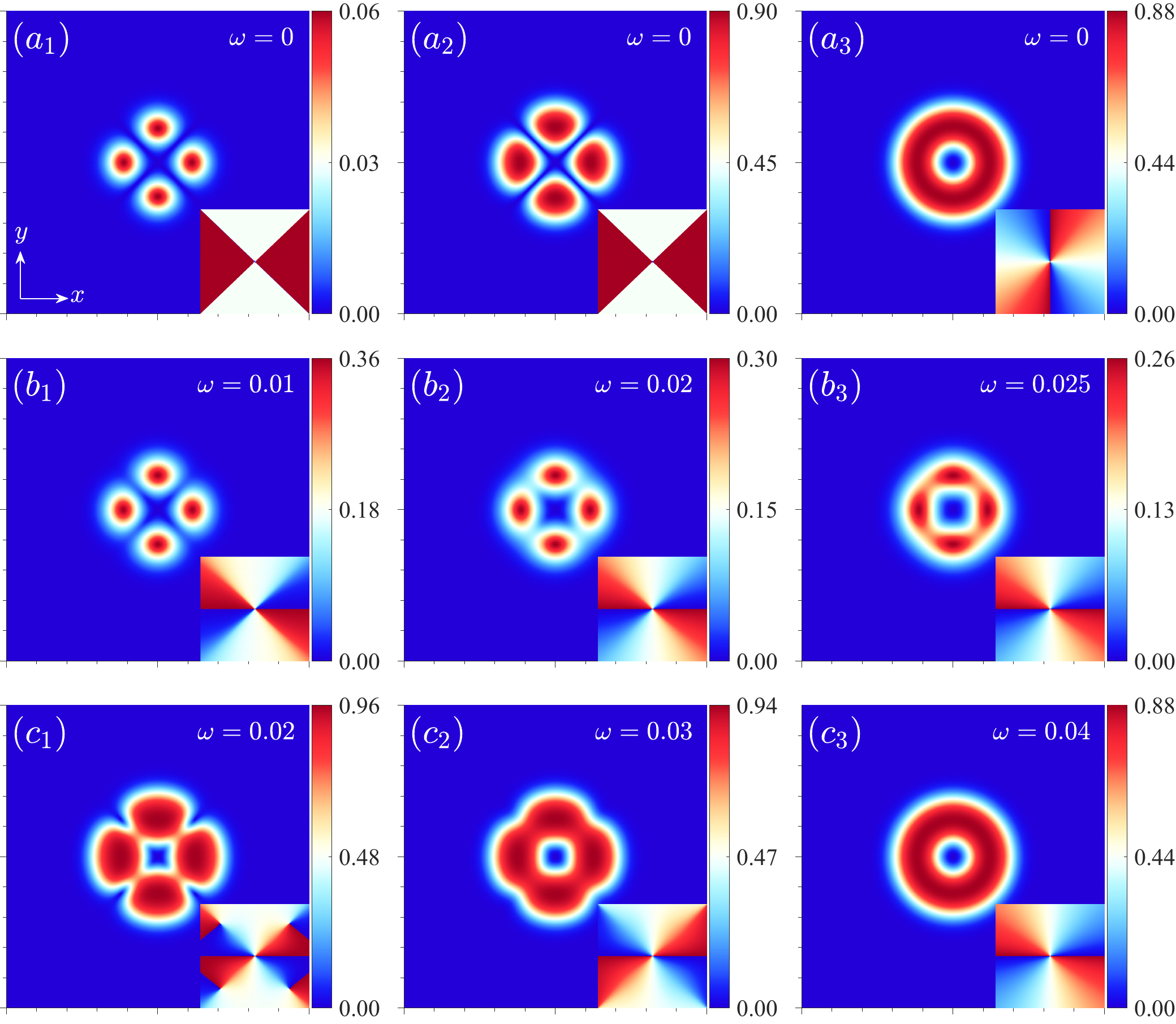} 
\caption{($a_{1}-a_{3}$): Profiles of QQDs and VQD with $S=2$ at $\protect%
\omega =0$, which are marked in Fig.~\protect\ref{fig:6}(a). The QQDs belong
to the lower and upper branches in ($a_{1}$) and ($a_{2}$), respectively. ($%
b_{1}-b_{3}$): The lower-branch QQDs which are marked in Fig.~\protect\ref%
{fig:6}(c) at $\protect\omega =0.01$, $\protect\omega =0.02$, and $\protect%
\omega =0.025$, respectively. $(c_{1})-(c_{3})$: The upper-branch QQDs and
VQD with $S=2$ which are marked in Fig.~\protect\ref{fig:6}(d) at $\protect%
\omega =0.02$, $\protect\omega =0.03$, and $\protect\omega =0.04$,
respectively. The chemical potential and rotation velocity are $\protect\mu %
=0.16,\protect\omega =0$ in ($a_{1}-a_{3}$) and $\protect\mu =0.26$ in ($%
b_{1}-b_{3}$), ($c_{1}-c_{3}$). }
\label{fig:5}
\end{figure}

The dependence of the norm of the QQDs and VQDs with $S=2$ on the chemical potential is displayed in Figs.~\ref{fig:6}(a,b). At $\omega =0$, the QQDs and VQDs with $S=2$ originate from similar linear eigenmodes at $\mu =0.418$. The rotating QQDs bifurcate from the VQDs with $S=2$, and merge with them again, eventually. At the lower branch, the norm decreases with the increase of $\omega $ [Fig.~\ref{fig:6}(c)]. The norm of the upper branch first increases and then decreases with the growth of $\omega $. This behavior of $N(\omega )$ for the QQDs is similar to that for DQDs reported above cf. Fig.~\ref{fig:2}(d).

\begin{figure}[tbph]
\centering
\includegraphics[width=0.9\linewidth]{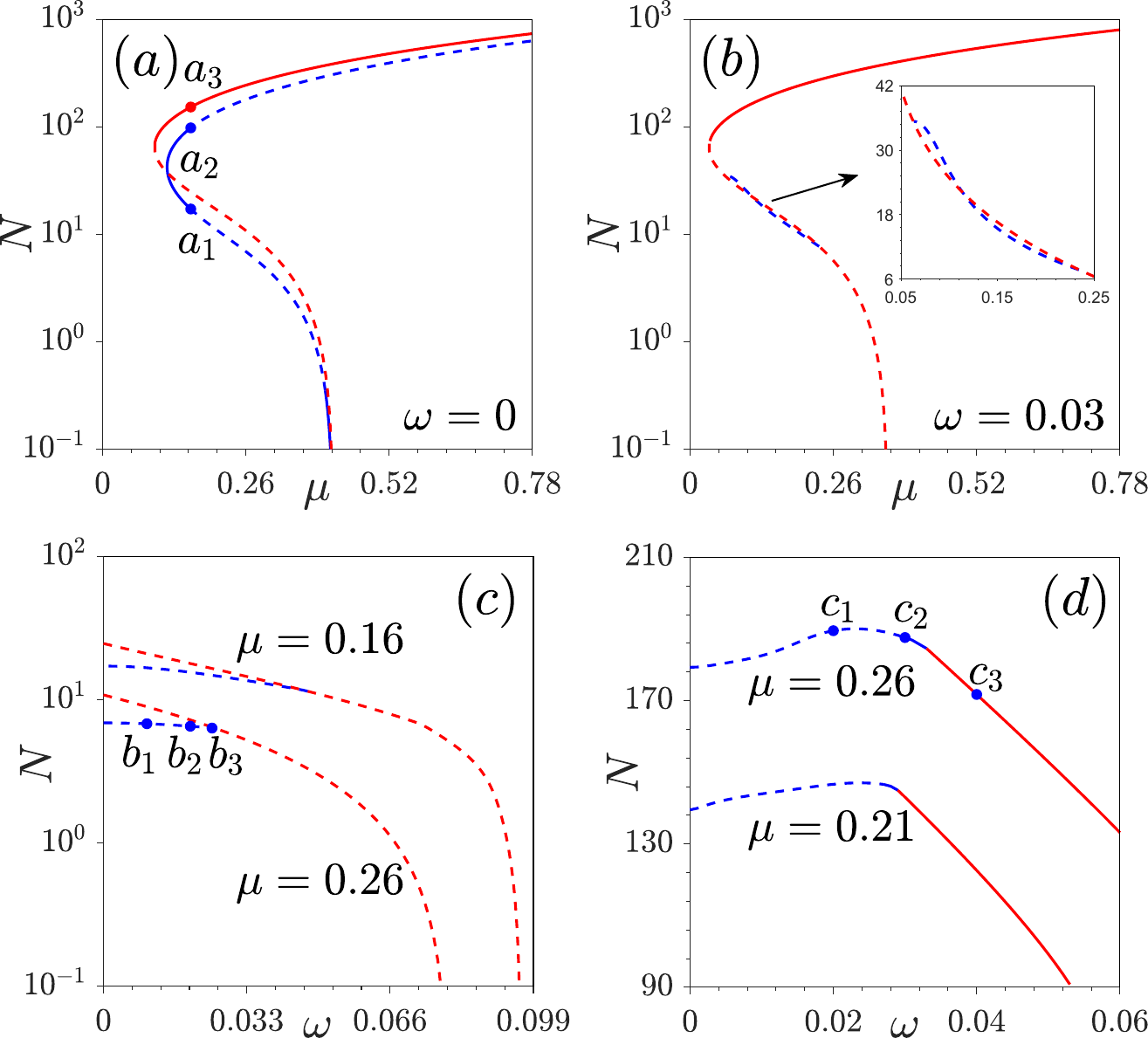}
\caption{$N(\protect\mu )$ curves for QQDs (blue) and VQDs with $S=2$ (red)
at (a) $\protect\omega =0$ and (b) $\protect\omega =0.03$. (c) $N(\protect%
\omega )$ for the lower-branch QQDs (blue) and VQDs with $S=2$ (red) at $%
\protect\mu =0.16$ and $\protect\mu =0.26$. (d) $N(\protect\omega )$ curves
for the upper-branch QQDs at $\protect\mu =0.21$ and $\protect\mu =0.26$. In
this panel, the red lines represent the VQDs emerging from the QQD states.
The solid and dashed segments designate unstable and stable subfamilies,
respectively. }
\label{fig:6}
\end{figure}

\begin{figure}[tbph]
\centering
\includegraphics[width=0.9\linewidth]{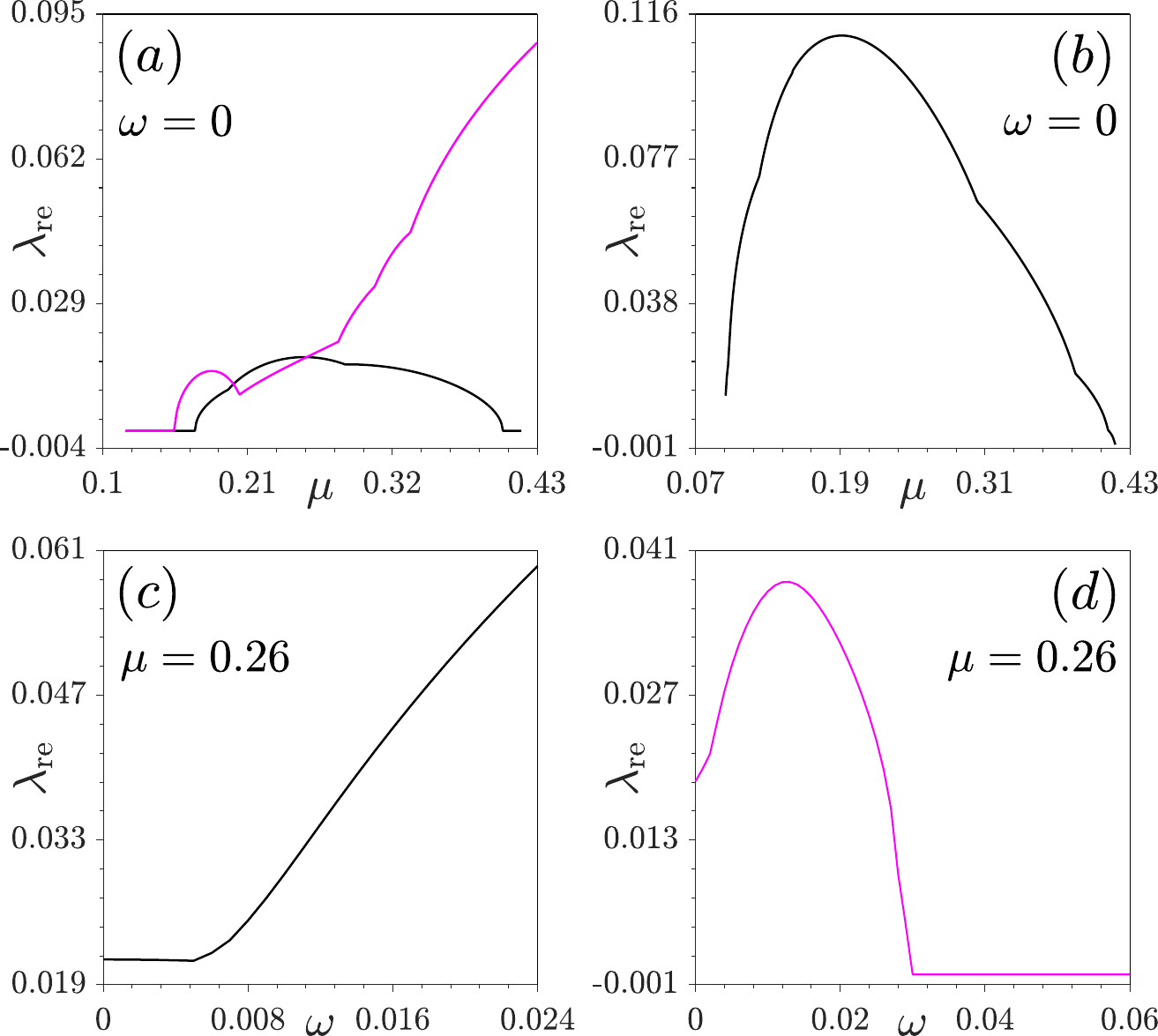}
\caption{Instability growth rate $\protect\lambda _{\mathrm{re}}$ vs. $%
\protect\mu $ for (a) the QQDs and (b) lower-branch VQDs with $S=2$ at $%
\protect\omega =0$. (c) and (d): $\protect\lambda _{\mathrm{re}}$ vs. $%
\protect\omega $ for the lower- and upper-branch QQDs, respectively, at $%
\protect\mu =0.26$. The lower and upper branches are represented by the
black and magenta curves, respectively. }
\label{fig:7}
\end{figure}

\begin{figure}[tbph!]
\centering
\includegraphics[width=\linewidth]{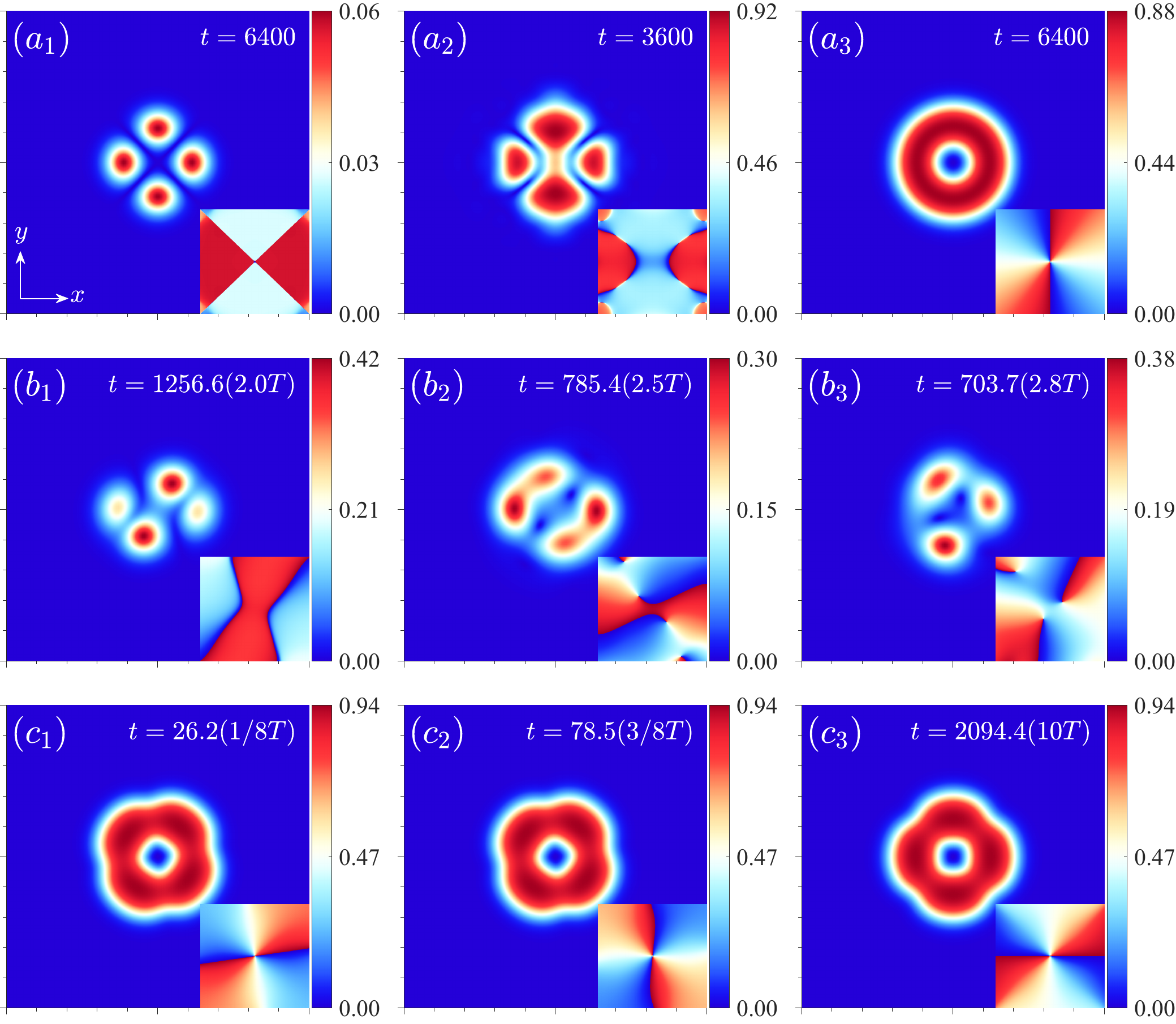} 
\caption{The stable evolution of ($a_{1}$) the lower-branch QQDs and ($a_{3}$%
) upper-branch vortices with $S=2$. ($a_{2}$) The unstable evolution of the
upper-branch QQDs. ($b_{1})-(b_{3}$): The unstable evolution of the
lower-branch QQDs at $\protect\omega =0.01$, $\protect\omega =0.02$, and $%
\protect\omega =0.025$, respectively. ($c_{1})-(c_{3}$) The stable evolution
of the upper-branch QQDs. The parameters are $\protect\omega =0$ in ($%
a_{1})-(a_{3}$) and $\protect\omega =0.03$ in ($c_{1})-(c_{3}$). The
chemical potential is $\protect\mu =0.16$ in ($a_{1})-(a_{3}$) and $\protect%
\mu =0.26$ in ($b_{1})-(b_{3}$), ($c_{1})-(c_{3}$). }
\label{fig:8}
\end{figure}

Compared to DQDs, the stability area of the lower-branch QQDs considerably shrinks at $\omega =0$, while it remains nearly the same for the upper-branch QQDs, see Fig.~\ref{fig:7}(a). For the VQDs with $S=2$, the
stability is similar to that for the vortices with $S=1$: as seen in Fig.~\ref{fig:7}(b), the lower-branch VQDs are completely unstable, and the upper-branch ones are stable in the entire existence domain. Further, the dependence of the instability growth rate on $\omega $ shows that the lower-branch QQDs are completely unstable [Fig.~\ref{fig:7}(c)], while the upper-branch ones are stable in a large domain, see Fig.~\ref{fig:7}(d).

The stable and unstable evolution of QQDs and VQDs with $S=2$, produced by direct simulations of Eq.~\ref{eq:refname2}, is illustrated in Fig.~\ref{fig:8}. Since the instability growth rate is small, the unstable QQDs survive for a long time at $\omega =0$, see Fig.~\ref{fig:8}($a_{2}$), while the rotating unstable QQDs break rapidly, see Figs.~\ref{fig:8}($b_{1}-b_{3}$). On the other hand, the stable QQDs demonstrate persistent rotation over long times in Figs.~\ref{fig:8}($c_{1}-c_{3}$).

\section{CONCLUSIONS}

We have predicted a new type of rotating QDs (quantum droplets) in binary BEC. These states connect the families of DQDs, QQDs, and VQDs (dipole, quadrupole, and vortex QDs, respectively). The interplay of the
LHY-corrected nonlinearity and annular trapping potential allows the existence of the stable rotating DQDs and QQDs, which bifurcate from the stable VQDs with winding numbers $S=1$ and $2$, respectively. With the
increase of the rotation frequencies, they spread out in the azimuthal direction and eventually fuse back into the VQDs. It is relevant to note that, in addition to these features predicted in BEC, similar ones are expected in models with competing nonlinearities, which occur in nonlinear optics \cite{reyna_high-order_2017}. Thus, our findings suggest a method for the creation of rotating quantum droplets and similar optical modes in the experiment. As an extension of the analysis, it may be relevant to consider the existence and stability of quiescent and rotating multipole necklace patterns, cf. Refs. \cite{kartashov_metastability_2019} and \cite{dong_multipole_2023}. 
\bigskip

\begin{acknowledgments}

This work is supported by the National Natural Science Foundation of China (project No. 61975130), 
Guangdong Basic and Applied Basic Research Foundation(project No. 2021A1515010084), and Israel Science Foundation (grant No. 1695/22).

\end{acknowledgments}

		
%

\end{document}